\documentclass[prd,preprint,a4paper,nofootinbib,showpacs,amsfonts,amssymb]{revtex4}
\usepackage{amsmath}
\usepackage{bm}
\usepackage{enumerate}
\usepackage{amssymb}

\begin{document}

\preprint{hep-th/0511157}

\title{\hspace{1pt}\\ \hspace{1pt}\\Canonical differential geometry of string backgrounds}
\author{Frederic P. Schuller}
\email{fschuller@perimeterinstitute.ca}
\affiliation{Perimeter Institute for Theoretical Physics, 31 Caroline Street N, Waterloo N2L 2Y5, Canada}
\author{Mattias N.\,R. Wohlfarth}
\email{mattias.wohlfarth@desy.de}
\affiliation{{II.} Institut f\"ur Theoretische Physik, Universit\"at Hamburg, Luruper Chaussee 149, 22761 Hamburg, Germany}

\begin{abstract}
String backgrounds and D-branes do not possess the structure of Lorentzian manifolds, but that of manifolds with area metric. Area metric geometry is a true generalization of metric geometry, which in particular may accommodate a $B$-field. While an area metric does not determine a connection, we identify the appropriate differential geometric structure which is of relevance for the minimal surface equation in such a generalized geometry. In particular the notion of a derivative action of areas on areas emerges naturally. Area metric geometry provides new tools in differential geometry, which promise to play a role in the description of gravitational dynamics on D-branes.
\end{abstract}

\maketitle

\numberwithin{paragraph}{section}

\section{Introduction}
Interest in a geometric understanding of generalized backgrounds in string theory has recently been fueled by Hitchin's proposal of a unified description of the spacetime metric and the Neveu-Schwarz two-form~\cite{Hitchin:2004ut}. This approach is of intrinsic mathematical appeal and has proven  valuable in compactifications of string theory on generalized complex manifolds~\cite{Grana:2004bg,Grana:2005ny}, and in studying D-branes and mirror symmetry \cite{Gualtieri:2004,Zucchini:2005rh,Koerber:2005qi,Fidanza:2003zi}. However, the basic premise of string theory, namely the replacement of point particles by strings, suggests an alternative geometric picture: manifolds equipped with an area measure. These present a true generalization of Lorentzian manifolds because only some area measures can be induced from a metric. In particular both a metric and a $B$-field may contribute to an area measure, as we will show, but the geometry of area metric manifolds is even more general.

This generality comes at a price; while the tangents to worldlines constitute vector spaces, the tangent areas of string worldsheets merely form a polynomial subspace, a so-called variety, of the vector space of antisymmetric two-tensors (which are often termed bivectors in the literature and play an important role in many areas of general relativity, for instance in the Petrov classification of the Weyl curvature \cite{Hall}). Tensors, being linear maps between vector spaces, are incompatible with the non-linear structure of a variety. This simple fact obstructs a routine construction of covariant derivatives and associated tensors from the area metric. In this paper, we develop the geometry of area manifolds aiming at the identification of relevant structures. We show that an area metric determines only part of a connection, namely what we term a symmetric pre-connection. The symmetric pre-connection in turn determines a derivative action of areas on areas which is 
sufficient to rewrite the string equation of motion, or the stationary surface equation, in a concise coordinate-independent form. This attaches particular relevance to the symmetric pre-connection because stationary surfaces probe the structure of area metric manifolds the same way geodesics probe the structure of metric manifolds.

The area geometry developed in this paper surfaces naturally in the context of string theory. We show that the Polyakov action for the fundamental string in a target space with non-vanishing Neveu-Schwarz two-form is classically equivalent to a surface area functional on an area metric manifold. The area measure on this manifold is the sum of the separate area measures induced from the metric and the two-form. Moreover, the area metric manifold structure reflects in the effective action for D-branes: these cannot be regarded as embedded Lorentzian manifolds, but find a neat geometric interpretation as area metric manifolds; the area metric in this case is a particular combination of the metric, the $B$-field and the $U(1)$ gauge field on the brane. Thus area geometry is a geometric way, alternative to Hitchin's, of understanding generalized geometries in string theory.

The construction of differential operators relevant for the geometry of areas in this paper is completely canonical: the area metric is treated as a fundamental structure on the manifold, which contrasts our recent work in \cite{Schuller:2005yt}, where area metric geometry was discussed as a multi-metric geometry by employing a particular Gilkey decomposition \cite{Gilkey:2001,Fiedler:2003} of the area metric. The identification of canonical differential geometric structures in the present paper sets the stage for the search for area geometric invariants which, in the light of our findings, promise to be of relevance for the description of gravity on D-branes \cite{Corley:2001hg,Ardalan:2002qt,Fotopoulos:2002wy,Cheung:2004sa}.

As a mathematical application we briefly consider the hierarchical structure of a manifold both with metric and independent area metric, which is a natural extension of metric manifolds from the point of view of category theory, and has been considered analogously in the context of higher gauge theories \cite{Pfeiffer:2003je,Girelli:2003ev,Baez:2004in}. While symmetric pre-connections cannot be used to define a covariant derivative, we show that any symmetric pre-connection can be extended to a proper connection on the antisymmetric two-tensors (into which the variety of areas is embedded), which then allows the construction of tensor invariants in standard fashion. This extension requires an anti-symmetric pre-connection which cannot be constructed from an area metric, but interestingly from a standard metric. The resulting  connection has the desirable property that areas are parallely transported into areas.

The organization of this paper is as follows. We define area metric manifolds in section~\ref{areaman}, and then proceed with the axiomatic definition of pre-connections in section \ref{preco} where we further show how a symmetric and an antisymmetric pre-connection may be combined to build a connection on the antisymmetric two-tensors. A canonical area derivative, completely defined in terms of the symmetric pre-connection, is introduced in section \ref{secareaderiv}. This area derivative is used in the following section \ref{surfaces} to rewrite the string equation of motion, which is discussed in full generality as a stationary surface on area metric manifolds. The application of area manifolds in understanding generalized string backgrounds and the identification of D-branes as area metric manifolds is made in section~\ref{DB}. With the help of both a symmetric and an antisymmetric pre-connection we construct tensors on manifolds equipped with a metric and an independent
area metric in section~\ref{secMgG}. We conclude with a discussion in section~\ref{Conc}.

\section{Area metric manifolds}\label{areaman}
We consider smooth $d$-dimensional manifolds $M$ equipped with a fourth rank covariant tensor $G$ with the following symmetries
\begin{equation}
  G(X,Y,A,B) = - G(Y,X,A,B) = G(A,B,X,Y)\,.
\end{equation}
Via linear extension the map $G$ naturally provides us with a linear map from the space of antisymmetric contravariant two-tensors $\bigwedge^2TM$ to its dual,
\begin{equation}
  G: \textstyle{\bigwedge^2} \,TM \longrightarrow (\textstyle{\bigwedge}^2\,TM)^*\,.
\end{equation}
In case the inverse $G^{-1}: (\bigwedge^2TM)^* \longrightarrow \bigwedge^2TM$ of the above map exists everywhere on $M$, we call $G$ an area metric and $(M,G)$ an area metric manifold. Note that the area metric $G$ may be uniquely decomposed as the sum of an algebraic curvature tensor and a four-form, which are irreducible under the local frame group $SL(d,\mathbb{R})$.

 Any metric manifold $(M,g)$ is an area metric manifold $(M,G_g)$, by virtue of $G_g(X,Y,A,B)=g(X,A)g(Y,B) - g(X,B)g(Y,A)$, which is readily seen to be an area metric. But not every area metric is induced by a single metric, but rather by a finite collection of metrics~$\{g^{(1)}, \dots,g^{(N)}\}$ via a decomposition theorem for algebraic curvature maps due to Gilkey~\cite{Gilkey:2001}. It follows from this theorem that any area metric can be decomposed as
\begin{equation}
  G =F+ \sum_{i=1}^N \sigma_{(i)}\, G_{g^{(i)}}\,,
\end{equation}
with signs $\sigma_{(i)} = \pm 1$ and a four-form $F$. Unfortunately, the decomposition is far from unique, and only a redefinition of area metric geometry as multi-metric geometry (by picking a particular decomposition), as in \cite{Schuller:2005yt}, allows to base the construction of curvature invariants on the constituent metrics $g^{(i)}$. In this paper we therefore do not consider any such decomposition.

It is not obvious how to construct some curvature tensor associated with an area metric~$G$ in a decomposition-independent fashion. This difficulty essentially roots in the fact that oriented areas, unlike vectors, do not constitute a linear space. More precisely, we observe that any area metric $G$ may indeed be used to consistently assign a surface area to any given oriented area $X\wedge Y$, where $X$ and $Y$ are two vectors in the same tangent space. Clearly, $X\wedge Y$ is an element of the vector space $\bigwedge^2TM$, but a generic antisymmetric two-tensor $\Omega \in \bigwedge^2TM$ may only be written as the exterior product of two vectors if $\Omega \wedge \Omega = 0$. Such elements of $\bigwedge^2TM$ are called simple and constitute the space of oriented areas $A^2TM$. As the simplicity condition is polynomial, $A^2TM$ is an affine variety embedded into the vector space $\bigwedge^2TM$:
\begin{equation}\label{embedding}
   A^2TM=\left\{\Omega\in\textstyle{\bigwedge}^2TM\,|\,\Omega\wedge\Omega\right\}.
\end{equation}
Strictly speaking, the area metric $G$ only ought to act on the variety $A^2TM$, which already renders $G$ non-tensorial. Indeed, it is not possible (without resorting to a particular Gilkey decomposition) to construct an affine connection from $G$, let alone tensors; not even on the embedding space $\bigwedge^2TM \supset A^2TM$.



\section{Pre-connections}\label{preco}
In this section we define symmetric and antisymmetric pre-connections on an area metric manifold $(M,G)$ into which any $\bigwedge^2TM$ connection uniquely decomposes. The following sections \ref{secareaderiv} and \ref{surfaces} then will show that symmetric pre-connections are of central importance for area geometry.

A symmetric pre-connection $D_\cdot(\cdot,\cdot)$ on an area metric manifold $(M,G)$ is a map sending a vector $X$ and two sections $\Omega,\Sigma$ of the bundle of antisymmetric two-tensors $\bigwedge^2TM$ to the function $D_X(\Omega,\Sigma)\in C^\infty (M)$ which satisfies the following properties
\begin{subequations}
\begin{eqnarray}
D_X(\Omega,\Sigma) &=& D_X(\Sigma,\Omega)\,,\\
D_{X+fY}(\Omega,\Sigma) &=& D_X(\Omega,\Sigma) + f D_Y(\Omega,\Sigma)\,,\\
D_X(\Omega,\Sigma+\Phi) &=& D_X(\Omega,\Sigma) + D_X(\Omega,\Phi)\,,\\
D_X(\Omega,f\Sigma) &=& fD_X(\Omega,\Sigma)+(Xf)G(\Omega,\Sigma)\,.
\end{eqnarray}
\end{subequations}
This means the symmetric pre-connection is $C^\infty (M)$-linear in $X$ but only $\mathbb{R}$-linear in~$\Omega$ and~$\Sigma$. The Leibniz rule in the last line holds identically for $D_X(f\Omega,\Sigma)$ because of the symmetry. A symmetric pre-connection is determined by provision of the coefficients $\Theta_{a_1a_2b_1b_2c}=D_c(e_{a_1}\wedge e_{a_2},e_{b_1}\wedge e_{b_2})$ which consequently appear in the coordinate expansion of the function $D_X(\Omega,\Sigma)$:
\begin{equation}\label{cospc}
D_X(\Omega,\Sigma)=X^c\partial_c(\Omega^{a_1a_2}\Sigma^{b_1b_2})G_{a_1a_2b_1b_2}+X^c\Theta_{a_1a_2b_1b_2c}\Omega^{a_1a_2}\Sigma^{b_1b_2}\,.
\end{equation}
Our conventions for the coordinate representations of objects and operations are explained in the  appendix. Under a coordinate change with the matrix $\phi^{a'}_a=\partial x^{a'}/\partial x^a$ the coefficients~$\Theta$ transform according to
\begin{equation}\label{trasym}
\Theta_{a_1a_2b_1b_2c}=\phi^{a'_1}_{a_1}\phi^{a'_2}_{a_2}\phi^{b'_1}_{b_1}\phi^{b'_2}_{b_2}\phi^{c'}_c\Theta_{a'_1a'_2b'_1b'_2c'}+\partial_c\left(\phi^{a'_1}_{a_1}\phi^{a'_2}_{a_2}\phi^{b'_1}_{b_1}\phi^{b'_2}_{b_2}\right)G_{a'_1a'_2b'_1b'_2}\,.
\end{equation}
We will find that symmetric pre-connections are at the heart of area metric geometry. In order to illuminate their relation to affine geometry we also introduce antisymmetric pre-connections and show that both are needed to compose connections.

An antisymmetric pre-connection $D_\cdot[\cdot,\cdot]$ on an area metric manifold $(M,G)$ is a map sending a vector $X$ and two sections $\Omega,\Sigma$ of the bundle of antisymmetric two-tensors $\bigwedge^2TM$ to the function $D_X[\Omega,\Sigma]\in C^\infty (M)$ which satisfies the following properties
\begin{subequations}
\begin{eqnarray}
D_X[\Omega,\Sigma] &=& -D_X[\Sigma,\Omega]\,,\\
D_{X+fY}[\Omega,\Sigma] &=& D_X[\Omega,\Sigma] + f D_Y[\Omega,\Sigma]\,,\\
D_X[\Omega,\Sigma+\Phi] &=& D_X[\Omega,\Sigma] + D_X[\Omega,\Phi]\,,\\
D_X[\Omega,f\Sigma] &=& f D_X[\Omega,\Sigma]-(Xf)G(\Omega,\Sigma)\,.
\end{eqnarray}
\end{subequations}
Note that the antisymmetry implies that
\begin{equation}
D_X[f\Omega,\Sigma] = f D_X[\Omega,\Sigma]+(Xf)G(\Omega,\Sigma)\,.
\end{equation}
The antisymmetric pre-connection is $C^\infty (M)$-linear in $X$ and $\mathbb{R}$-linear in $\Omega$ and $\Sigma$. It is determined by coefficients $\Xi_{a_1a_2b_1b_2c}=D_c[e_{a_1}\wedge e_{a_2},e_{b_1}\wedge e_{b_2}]$ which under a coordinate change transform according to
\begin{eqnarray}\label{apretrans}
\Xi_{a_1a_2b_1b_2c} &=& \phi^{a'_1}_{a_1}\phi^{a'_2}_{a_2}\phi^{b'_1}_{b_1}\phi^{b'_2}_{b_2}\phi^{c'}_c\Xi_{a'_1a'_2b'_1b'_2c'}\nonumber\\
& & +\left[\partial_c\left(\phi^{a'_1}_{a_1}\phi^{a'_2}_{a_2}\right)\phi^{b'_1}_{b_1}\phi^{b'_2}_{b_2}-\phi^{a'_1}_{a_1}\phi^{a'_2}_{a_2}\partial_c\left(\phi^{b'_1}_{b_1}\phi^{b'_2}_{b_2}\right)\right]G_{a'_1a'_2b'_1b'_2}
\,.
\end{eqnarray}
The coordinate expansion of the function $D_X[\Omega,\Sigma]$ reads
\begin{equation}
D_X[\Omega,\Sigma]=X^c\left(\partial_c\Omega^{a_1a_2}\Sigma^{b_1b_2}-\Omega^{a_1a_2}\partial_c\Sigma^{b_1b_2} \right)G_{a_1a_2b_1b_2}+X^c\Xi_{a_1a_2b_1b_2c}\Omega^{a_1a_2}\Sigma^{b_1b_2}\,.
\end{equation}
Due to the antisymmetry the two terms in bracket cannot be combined using the product rule, in contrast to the symmetric case of expression (\ref{cospc}) above.

We now show that on an area metric manifold any $\bigwedge^2TM$ connection uniquely decomposes into a symmetric and an antisymmetric pre-connection. An important corollary of this insight is the existence of pre-connections. Consider the following sum of a symmetric and an antisymmetric pre-connection:
\begin{equation}
  D_X(\Omega,\Sigma)+D_X[\Omega,\Sigma]\,.
\end{equation}
Due to the different symmetry properties of the pre-connections, the above sum is now $C^\infty (M)$-linear in $\Sigma$, which puts us into the position to define a connection $\nabla$ on the bundle~$\bigwedge^2TM$ by letting
\begin{equation}\label{connL}
\nabla_X\Omega=\frac{1}{2}G^{-1}(D_X(\Omega,\cdot)+ D_X[\Omega,\cdot],\cdot)\,.
\end{equation}
Indeed, the vector entry $X$ is $C^\infty (M)$-linear, and the entry $\Omega$ is $\mathbb{R}$-linear. Under smooth rescalings of $\Omega$ by a function $f$ on the manifold we also find the expected Leibniz rule
\begin{equation}
\nabla_X(f\Omega)=f\nabla_X\Omega+(Xf)\Omega\,.
\end{equation}
The coefficients $\Gamma^{c_1c_2}{}_{a_1a_2b}$ of the $\bigwedge^2TM$ bundle connection are easily expressed in terms of those of the pre-connections by
\begin{equation}
\Gamma^{c_1c_2}{}_{a_1a_2b}=\frac{1}{2}G^{c_1c_2p_1p_2}\left(\Theta_{a_1a_2p_1p_2b}+\Xi_{a_1a_2p_1p_2b}\right).
\end{equation}
Conversely, by the invertibility of the area metric $G$ this shows that on an area metric manifold any $\bigwedge^2TM$ connection can be uniquely decomposed into a symmetric and an antisymmetric pre-connection.

There is a natural compatibility criterion for $\bigwedge^2TM$ connections with respect to the area metric $G$ which fully determines the symmetric pre-connection. Let $\nabla$ be an arbitrary connection on the $\bigwedge^2TM$ bundle over an area metric manifold $(M,G)$. The connection $\nabla$ is called area metric compatible if the area metric is preserved under parallel transport, i.e., if
\begin{equation}
  \nabla_X G = 0\,.
\end{equation}
While this requirement is formulated in terms of a covariant derivative, it does only present a condition on the symmetric pre-connection $D_\cdot(\cdot,\cdot)$ because
\begin{equation}
  (\nabla_X G)(\Omega,\Sigma) = X G(\Omega,\Sigma) - D_X(\Omega,\Sigma)\,.
\end{equation}
Importantly, it follows that area metric compatibility uniquely determines the symmetric pre-connection coefficients $\Theta$ as
\begin{equation}\label{ampconn}
   \Theta_{a_1 a_2 b_1 b_2 c} = \partial_c G_{a_1 a_2 b_1 b_2}\,.
\end{equation}

A further geometric implication of area metric compatibility is that areas are parallely transported into areas on even-dimensional area metric manifolds. To see this consider the volume form $\omega$ on $(M,G)$ with components
\begin{equation}\label{volume}
\omega_{a_1\dots a_d} = ((-1)^{d-1}\det G)^{1/(2d-2)}\varepsilon_{a_1\dots a_d}\,,
\end{equation}
where $\varepsilon$ is the totally antisymmetric density, and the determinant is calculated for ${G:\bigwedge^2TM\rightarrow (\bigwedge^2TM)^*}$. Area metric compatibility implies
\begin{equation}
\Gamma^{mn}{}_{mnb}=\frac{1}{4}G^{mnpq}\Theta_{mnpqb}=\frac{1}{4}G^{mnpq}\partial_bG_{mnpq}\,,
\end{equation}
which is equivalent to $\nabla_X\omega=0$ as shown in \cite{Schuller:2005yt}. This in turn implies that the simplicity condition $\Omega\wedge\Omega=0$ (for $\Omega$ to be an area in $A^2TM$), which may be written ${\omega(\Omega,\Omega,*,\dots,*)=0}$, is preserved under parallel transport.
Since area metric compatibility preserves the inner product under parallel transport, it also preserves the normalization of any $\bigwedge^2TM$ basis $\{e_A\}$ which one may choose as $G(e_A,e_B)=\widetilde\eta_{AB}$, with the pseudo-Riemannian area metric $\widetilde\eta = \textrm{diag}(-1,\dots,-1,1,\dots,1)$ of signature $\left(d-1,\left({}^{d-1}_{\;\;2}\right)\right)$. 

We may thus show that the structure group $GL(d)$, which acts in a $\left({}^d_2\right)$-dimensional representation on the $\bigwedge^2TM$ bundle, is reduced by the introduction of an area metric. Generically, the action of a transformation $P$ in $GL(d,\mathbb{R})$ on $\Omega$ in $\bigwedge^2TM$ is given by
\begin{equation}\label{actsgroup}
\left(P\Omega\right)^{a_1a_2}=2P^{a_1}{}_{b_1}P^{a_2}{}_{b_2}\Omega^{b_1b_2}\,. 
\end{equation}
Now consider a transformation $P$ in $GL(d)$ which respects the orthonormality of the basis. Using~(\ref{actsgroup}), the required condition $G(Pe_A,Pe_B)=G(e_A,e_B)$ is then equivalent to
\begin{equation}
\widetilde\eta_{a_1a_2b_1b_2}=4\widetilde\eta_{c_1c_2d_1d_2}P^{c_1}{}_{a_1}P^{c_2}{}_{a_2}P^{d_1}{}_{b_1}P^{d_2}{}_{b_2}\,.
\end{equation} 
In contrast to the general case, the area metric $\widetilde\eta$ can always be induced from a single metric, namely the Minkowski metric $\eta$, since $\widetilde\eta_{a_1a_2b_1b_2}=\eta_{a_1b_1}\eta_{a_2b_2}-\eta_{a_1b_2}\eta_{a_2b_1}$. The above condition can thus be rewritten as
 \begin{equation}\label{fullcond}
Q^{[a}{}_{c}Q^{b]}{}_{d}=\delta^{[a}_c\delta^{b]}_d
\end{equation} 
for $Q=\eta^{-1}P^t\eta P$. Taking a simple trace yields $(\mathrm{Tr}\,Q) Q - Q^2 = (d-1) \delta$, where $Q$ can be brought to Jordan normal form (over the complex numbers) and so the traced condition must be satisfied for each Jordan block. If any one of the blocks were non-diagonal the traced condition could not hold, as $Q^2$ then would have a second superdiagonal. Hence the Jordan normal form of Q is diagonal, and the full condition (\ref{fullcond}), also transformed to Jordan normal form, can be written in terms of eigenvalues $q_{(a)}$ of $Q$ as $(q_{(a)}q_{(b)}-1)\delta^{[a}_c\delta^{b]}_d=0$. It follows that $q_{(a)}q_{(b)}=1$ for all $a\neq b$, so that, in dimension $d>2$, all eigenvalues of Q are either $+1$ or $-1$. Hence $Q=\pm\delta$, or $P^t\eta P=\pm \eta$. For the lower sign the invertible transformation $P$ maps spacelike vectors to timelike ones and vice versa. For a real transformation $P$ this is only possible for metrics $\eta$ of signature $(s,s)$ which are not Lorentzian for $d>2$. For the upper sign, $P$ is a Lorentz transformation. In our case therefore, the structure group is reduced as in the case of standard metric compatibility $GL(d)\rightarrow SO(1,d-1)$. This is compatible with the notion of the Lorentzian structure of an area metric manifold which is still apparent from the cones formed by the null planes. Note that in contrast to metric geometry, no condition such as vanishing torsion is needed to fully determine the coefficients of the symmetric pre-connection.


\section{Canonical area derivative}\label{secareaderiv}
We have seen that on an area metric manifold $(M,G)$, a connection uniquely decomposes into a symmetric and an antisymmetric pre-connection, and that the symmetric pre-connection is uniquely determined by the requirement of covariant constancy of the area metric. In section \ref{surfaces} we will demonstrate that the equation for stationary surfaces can be written in terms of a derivative action of areas on areas, whose definition solely depends on the symmetric pre-connection. For the sake of clarity we define this derivative action before relating it to the stationary surface equation.

More precisely, we now construct a $T^*M$-valued derivative $\mathcal{D}_\Sigma\Omega$ of an area ${\Sigma=X\wedge Y}$ in $A^2T_pM$ acting on a section $\Omega$ of the bundle $\bigwedge^2TM$ of antisymmetric two-tensors. The definition will only involve the symmetric pre-connection: for any vector $Z$ we define
\begin{eqnarray}
\mathcal{D}_\Sigma\Omega (Z)=D_X(\Omega,Y\wedge Z)-\frac{1}{2}G(\Omega,\mathcal{L}_XY\wedge Z)+\textrm{terms cyclic in } X,Y,Z\,.
\end{eqnarray}
This is well-defined as the right-hand side is invariant under $SL(2,\mathbb{R})$-transformations of the representatives $X,Y$ for $\Sigma$. Checking this requires some amount of algebra and the crucial fact that the derivative of the unit determinant of the transformation vanishes.
One easily sees that for a function $f$ and a vector $W$
\begin{equation}
\mathcal{D}_\Sigma\Omega(fZ+W) = f\mathcal{D}_\Sigma\Omega(Z)+\mathcal{D}_\Sigma\Omega(W)\,,
\end{equation}
so that $\mathcal{D}_\Sigma\Omega$ is indeed a one-form. We cannot expect additivity in $\Sigma$, as no addition of generic elements can be defined on $A^2TM$, except in case the two summand areas intersect. The derivative is, however, homogeneous in $\Sigma$,
\begin{equation}
\mathcal{D}_{f\Sigma}\Omega=f\mathcal{D}_\Sigma\Omega\,.
\end{equation}
The area derivative $\mathcal{D}_\Sigma\Omega$ displays the additivity property for sections $\Omega,\Phi$ of $\bigwedge^2TM$; but an inhomogeneous term is generated under rescalings of $\Omega$ by a function $f$:\begin{subequations}
\begin{eqnarray}
\mathcal{D}_\Sigma (\Omega+\Phi) &=& \mathcal{D}_\Sigma\Omega+\mathcal{D}_\Sigma\Phi\,,\\
\mathcal{D}_\Sigma(f\Omega)(Z) &=& f\mathcal{D}_\Sigma\Omega(Z)+3 G(\Omega,(\Sigma\wedge Z)\llcorner df)\,.
\end{eqnarray}
\end{subequations}
The occurrence of the area metric $G$ in the rescaling property derives from the fact that it is used in the definition of the symmetric pre-connection.
For the coordinate expansion of $\mathcal{D}_\Sigma\Omega(Z)$ we find
\begin{equation}
\mathcal{D}_\Sigma\Omega(Z)=\frac{3}{4}\Sigma^{[ad}Z^{e]}\left(G_{bcad}\partial_e\Omega^{bc}+\Omega^{bc}\Theta_{bcade}\right),
\end{equation}
which again explicitly displays the dependence of the area derivative on the symmetric pre-connection only. For the area metric compatible symmetric pre-connection defined by the coefficients (\ref{ampconn}) this can be rewritten as the following evaluation of a three-form:
\begin{equation}\label{DO}
\mathcal{D}_\Sigma\Omega(Z)=d[G(\Omega,\cdot)](\Sigma\wedge Z)\,.
\end{equation}

\section{Stationary surfaces}\label{surfaces}
Stationary surfaces probe the geometry of area metric manifolds $(M,G)$; they are geometrically well-defined objects obtained by variation of the surface area functional defined by~$G$. We demonstrate that the corresponding Euler-Lagrange equation only depends on the symmetric pre-connection. In gauge-fixed form the equation requires the vanishing of~$\mathcal{D}_\Omega\Omega$ which is thus identified as the mean curvature one-form of the surface.

We begin by considering the Nambu-Goto action which measures the worldsheet area of a string moving in the target space manifold $M$. The embedding functions $x^a$ depend on the parametrization by the worldsheet coordinates $\sigma^\alpha=(\tau,\sigma)$. The classical form of the action is based on length measurement through a target space metric $g$ pulled-back to the worldsheet,
\begin{equation}\label{origstac}
S_\mathrm{NG}=\int d^2\sigma \sqrt{\det \partial_\alpha x^a \partial_\beta x^b g_{ab}(x)}\,.
\end{equation}
The determinant under the square root can be expanded, yielding ${(G_g)_{abcd}\dot x^a x'^b\dot x^c x'^d}$ with the induced area metric $(G_g)_{abcd}=g_{ac}g_{bd}-g_{ad}g_{bc}$.

In order to obtain the action that is our generalized starting point we now replace the induced area metric $G_g$ by a general area metric~$G$. Abbreviating $\Omega=\dot x\wedge x'$ with ${\Omega^{ab}=2\dot x^{[a}x'^{b]}}$, where dot and prime respectively denote differentiation with respect to $\tau$ and $\sigma$, we then have
\begin{equation}\label{stringaction}
S=\frac{1}{2}\int d^2\sigma \sqrt{G_{abcd}(x)\Omega^{ab}\Omega^{cd}}\,.
\end{equation}
Note that $\Omega=\dot x\wedge x'$ is a section of $A^2TM$. But not any section of $A^2TM$ is integrable, i.e., arises as the tangent area distribution of some surface. The necessary and sufficient criterion for integrability is the Frobenius criterion which can be concisely written
\begin{equation}\label{inte}
\{\Omega,\Omega\}=0\,.
\end{equation}
The anticommutator bracket maps two $\bigwedge^2TM$ sections to $\bigwedge^3TM$ and is defined by its action on one-forms \cite{Schuller:2005yt} as follows
\begin{equation}
\{\Omega,\Sigma\}\,\llcorner\,\omega=\frac{1}{3}\left(\mathcal{L}_{\Sigma\,\llcorner\,\omega}\Omega+\mathcal{L}_{\Omega\,\llcorner\,\omega}\Sigma-2\,\Sigma\,\llcorner\, d\omega\,\lrcorner\,\Omega-2\,\Omega\,\llcorner\, d\omega\,\lrcorner\,\Sigma   \right),
\end{equation}
where $\mathcal{L}$ denotes the Lie derivative and the conventions for the contraction symbols are explained in the appendix.

We vary the area metric string action (\ref{stringaction}) with respect to the worldsheet embedding functions $x^a(\sigma^\alpha)$ to obtain the minimal surface equation in terms of the area metric $G$ and its partial derivatives, and $\Omega=\dot x\wedge x'$. This yields
\begin{equation}\label{surface}
\Omega^{af}\partial_f \Omega^{cd}G_{abcd}+\Omega^{af}\Omega^{cd}\left(\partial_f G_{abcd}-\frac{1}{4}\partial_b G_{afcd}\right)=\frac{1}{2}\Omega^{af}\Omega^{cd}G_{abcd}\partial_f \ln \left(G_{pqrs}\Omega^{pq}\Omega^{rs}\right),
\end{equation}
if boundary conditions are chosen such that
\begin{equation}
\int d\sigma \frac{G_{abcd}\Omega^{cd}}{\sqrt{G_{pqrs}\Omega^{pq}\Omega^{rs}}}\delta x^{[a}x'^{b]}+\int d\tau \frac{G_{abcd}\Omega^{cd}}{\sqrt{G_{pqrs}\Omega^{pq}\Omega^{rs}}}\dot x^{[a} \delta x^{b]}=0\,.
\end{equation}
Here the first integral is evaluated at the boundary of the original $\tau$-range, and the second integral is evaluated at the boundary of the original $\sigma$-range.

The stationarity condition (\ref{surface}) may be written in the coordinate-free form
\begin{equation}
Z\,G(\Omega,\Omega)-2\,d[G(\Omega,\cdot)](\Omega\wedge Z)=-2\,G(\Omega,Z\wedge [\Omega\,\llcorner\,  d\ln G(\Omega,\Omega)])\,.
\end{equation}
It is now easily checked that this equation is reparametrization invariant under $\Omega\mapsto f\Omega$. Note that no reparametrization affects the integrability of an area distribution $\Omega$ in $A^2TM$ since $\{f\Omega,f\Omega\}=f^2\{\Omega,\Omega\}$. Choosing a parametrization with constant $G(\Omega,\Omega)$ removes the first term and the right hand side. We then see that the stationary surface equation can be written in terms of an area metric compatible symmetric pre-connection using equation~(\ref{DO}) as
\begin{equation}
\mathcal{D}_\Omega\Omega=0\,,
\end{equation}
at which point the central relevance of the symmetric pre-connection and area derivative for area metric geometry becomes manifest. The stationary surface equation for an embedded surface is thus written as the condition on its tangent area distribution $\Omega$ which is a section of $A^2TM$, if it is supplemented by the integrability condition (\ref{inte}).

The stationary surface equation requires the vanishing of the one-form $\mathcal{D}_\Omega\Omega$ which therefore presents a generalization of the mean curvature vector that vanishes for any two-dimensional minimal surface embedded into a Riemannian manifold.

\section{String backgrounds and D-branes}\label{DB}
Area metric geometry provides a natural framework for the discussion of classical strings. We show that the fundamental string action on a generalized background can be cast into the form of a surface action on an area metric manifold, and that this clarifies the equal roles played by $g$ and~$B$ in the formation of the area measure. Most interestingly, the area metric manifold structure is still apparent in the D-brane effective action. This suggests that gravity induced on D-branes could be given by area metric manifold dynamics.

Strings on a generalized background consisting of a metric $g$ and a Neveu-Schwarz two-form $B$ are governed by the Polyakov worldsheet action
\begin{equation}\label{Polya}
S_P=\frac{1}{2}\int d^2\sigma \sqrt{-\gamma}\left(\gamma^{\alpha\beta}+\epsilon^{\alpha\beta}\right)\partial_\alpha x^a\partial_\beta x^b \left(g_{ab}+B_{ab}\right),
\end{equation}
where $\gamma$ is the worldsheet metric and $\epsilon_{\alpha\beta}=\sqrt{-\gamma}\varepsilon_{\alpha\beta}$ the antisymmetric Levi-Civita tensor. This action can be rewritten in the following form,
\begin{equation}
S_P=\frac{1}{2}\int d^2\sigma\sqrt{-\left(1-\lambda^2\right)\det(\gamma-\lambda\epsilon)}\left(\gamma-\lambda\epsilon\right)^{-1\;\alpha\beta}\partial_\alpha x^a\partial_\beta x^b \left(g_{ab}+\lambda^{-1}B_{ab}\right),
\end{equation}
as can be seen by expanding $\det(\gamma-\lambda\epsilon)$ and $(\gamma-\lambda\epsilon)^{-1}$. Formally, this looks like a Polyakov action with worldsheet metric $\tilde\gamma=\gamma-\lambda\epsilon$, which however is non-symmetric and thus does not admit a straightforward geometric interpretation. The combination $\gamma-\lambda\epsilon$ can be eliminated from this action in standard fashion by substituting the algebraic `energy-momentum' constraint obtained by variation with respect to $\tilde\gamma$, leading to the classically equivalent Nambu-Goto type action
\begin{equation}\label{NGtypeaction}
S_\textrm{NG}=\int d^2\sigma \sqrt{-\left(1-\lambda^2\right)\det\;\partial_\alpha x^a\partial_\beta x^b (g_{ab}+\lambda^{-1}B_{ab})}\,.
\end{equation}

To see that this action has a clear interpretation within area metric geometry, we first define for any invertible map $m:TM\rightarrow T^*M$ an induced area metric $G_m:\bigwedge^2TM\rightarrow \bigwedge^2T^*M$ by
\begin{equation}
  G_m(X,Y,U,V) = \frac{1}{2}\Big[m(X,U)m(Y,V)-m(X,V)m(Y,U)+(X,Y)\leftrightarrow (U,V) \Big].
\end{equation}
Note that the explicit symmetrization is not needed in case $m$ is either purely symmetric or antisymmetric. Using this definition one immediately finds, with
 $\Omega=\dot x\wedge x'$,
\begin{equation}\label{areaNG}
S_\textrm{NG}=\int d^2\sigma\sqrt{\left(1-\lambda^2\right)G_{g+\lambda^{-1}B}(\Omega,\Omega)}\,.
\end{equation}
While there is no meaningful interpretation for the non-symmetric tensor $g+\lambda^{-1}B$ as a length measure, $G_{g+\lambda^{-1}B}$ is a meaningful symmetric area metric. The classical equivalence of (\ref{areaNG}) and the Polyakov action (\ref{Polya}) shows that the classical string background with $g$ and $B$ can be interpreted as an area metric manifold $(M,G_{g+\lambda^{-1}B})$ for any $\lambda$ with $\lambda^2\neq 0,1$. Hence there is a one-parameter family of area metric manifolds representing the geometry of string backgrounds with non-vanishing $B$-field. The simple fact
\begin{equation}
G_{g+\lambda^{-1}B}(\Omega,\Sigma)=G_{g}(\Omega,\Sigma)+G_{\lambda^{-1}B}(\Omega,\Sigma)
\end{equation}
further reveals that both $g$ and $B$ independently contribute to the total area measure, while none of them plays a distinguished role.

The area metric manifold structure $(M,G_{g+B})$ of the fundamental string action is reflected also in the string-induced Dirac-Born-Infeld action which appears as the effective action  for a gauge field $A$ induced by open strings ending on a D-brane $M$ {\cite{Fradkin:1985qd,Abouelsaood:1986gd,Bergshoeff:1987at,Leigh:1989jq}}:
\begin{equation}
  S_{\textrm{DBI}} = \int_M \sqrt{-\det(g + B+2\pi\alpha'F)}\,.
\end{equation}
This may be interpreted as the volume of the D-brane, but the volume form cannot be that of a Lorentzian manifold (because this would require a non-symmetric metric $g+B+2\pi\alpha'F$). An area metric, on the other hand, can absorb the modification by $B$ and by the gauge field. The Born-Infeld action, and thus the D-brane volume, can be written as the canonical volume (\ref{volume}) of an area metric manifold.

This becomes possible through the following determinant identity for any map $m:TM\rightarrow T^*M$, for which $m=m_s+m_a$ is the canonical split of $m$ into its symmetric and antisymmetric part:
\begin{equation}
  (\det m)^{d-1}=\det\left(G_{m_s}+\kappa_d(m)G_{m_a}\right),
\end{equation}
where the area metric determinant on the right hand side is calculated for the corresponding matrix in the $\bigwedge^2TM$ basis. Interestingly, a scaling function $\kappa_d(m)$ depending on the spacetime dimension and on the matrix $m$ appears in front of the area metric contribution induced from the antisymmetric $m_a$, as for the fundamental string ($\kappa\sim\lambda^{-2}$). While computer algebra shows the existence of this function for any dimension $2\leq d\leq 11\dots$, it is very difficult to obtain its analytical expression for larger $d$. For $d=2$ one simply finds constant $\kappa_2(m)=1$, for $d=3$ the expression becomes $\kappa_3(m)=1+\det (m_s^{-1}m)$.

Using the determinant identity, we now rewrite the Dirac-Born-Infeld action as
\begin{equation}
  S_{\textrm{DBI}} = \int_M \left((-1)^{d-1} \det \left(G_g+\kappa_d(g+B+2\pi\alpha'F)G_{B+2\pi\alpha'F}\right)\right)^{1/(2d-2)}\,.
\end{equation}
Hence D-branes are area metric manifolds, but not Lorentzian manifolds.

\section{Application to manifolds with metric and area metric}\label{secMgG}
We have identified the symmetric pre-connection and the associated area derivative operator as structures of central importance to area metric geometry. In particular we have shown that the area metric induced symmetric pre-connection determines the stationary surface equation.
It seems not possible, however, to construct a $\bigwedge^2TM$ connection from an area metric alone. This is because there is no antisymmetric pre-connection in terms of $G$ which could be used to complement the symmetric pre-connection.

In the search for geometric invariants of area metric manifolds one may now proceed along two different routes.  One possibility is the construction of invariants only involving the symmetric pre-connection, which is entirely in the spirit of area metric geometry, but complicated by the non-linear structure of area spaces. This approach is currently under investigation.
The second possibility, which we will briefly explore in this section, is to supply additional structure in order to extend the symmetric pre-connection to a connection.

More specifically, we study the hierarchical structure of a manifold $(M,g,G)$ with both metric and area metric. The existence of the standard metric $g$ immediately induces the metric compatible Levi-Civita connection $\nabla^\textrm{TM}$ on the tangent bundle which may now be used to define an antisymmetric pre-connection
\begin{equation}
D_X[\Omega,\Sigma]=G(\nabla^\textrm{TM}_X\Omega,\Sigma)-G(\Omega,\nabla^\textrm{TM}_X\Sigma)
\end{equation}
for any vector X and sections $\Omega,\,\Sigma$ of $\bigwedge^2TM$. In coefficients this reads
\begin{equation}
\Xi_{a_1a_2b_1b_2c}=2\Gamma^p{}_{[a_1|c}G{}_{p|a_2]b_1b_2}-(a_1a_2\leftrightarrow b_1b_2)\,,
\end{equation}
where $\nabla^\textrm{TM}_ae_b=\Gamma^c{}_{ba}e_c$ defines the coefficients of $\nabla^\textrm{TM}$. The extension of the area metric compatible symmetric pre-connection by the above antisymmetric pre-connection leads to a $\bigwedge^2TM$ connection as defined in (\ref{connL}),
\begin{equation}
\nabla_X\Omega=\nabla^\textrm{TM}_X\Omega+\frac{1}{2}G^{-1}((\nabla^\textrm{TM}_XG)(\Omega,\cdot),\cdot)\,,
\end{equation}
whose coefficients have the simple form
\begin{equation}\label{hjk}
\Gamma^{a_1a_2}{}_{b_1b_2c}=4\delta{}^{[a_1}_{[b_1}\Gamma{}^{a_2]}_{\quad b_2]c}+\frac{1}{4}G^{a_1a_2pq}\nabla^\textrm{TM}_cG_{pqb_1b_2}\,.
\end{equation}
The affine structure provided by this connection on $\bigwedge^2TM$ allows the definition of tensors in standard fashion. We immediately may define the curvature tensor associated to $\nabla$ as
\begin{equation}\label{curv}
R(X,Y)\Omega=\nabla_X\nabla_Y\Omega-\nabla_Y\nabla_X\Omega-\nabla_{[X,Y]}\Omega
\end{equation}
for vector fields $X,\,Y$ and sections $\Omega$ of $\bigwedge^2TM$. There is also a tensor of first derivative order, namely
\begin{equation}\label{Tten}
T(X,Y,Z)=\nabla_X(Y\wedge Z)-[X,Y]\wedge Z+\textrm{terms cyclic in }X,Y,Z\,.
\end{equation}
The components of these tensors are given by\begin{subequations}
\begin{eqnarray}
\!\!\!\!\!\!\!R^{c_1c_2}{}_{a_1a_2mn} &= & 4R^{[c_1}{}_{[a_1|mn|}\delta^{c_2]}_{a_2]}+\left(\nabla^\textrm{TM}_mX^{c_1c_2}{}_{a_1a_2n}+X^{c_1c_2}{}_{f_1f_2m}X^{f_1f_2}{}_{a_1a_2n}-(m\leftrightarrow n)\right)\!,\\
\!\!\!\!\!\!\!T^{c_1c_2}{}_{mnp} & = & 3X^{c_1c_2}{}_{[mnp]}\,,
\end{eqnarray}
\end{subequations}
where the Riemann tensor of $\nabla^\textrm{TM}$ appears. The tensor $X$ is the difference between the connections~$\nabla$ and~$\nabla^\textrm{TM}$ on $\bigwedge^2TM$, given by the second term on the right hand side of equation~(\ref{hjk}). In case the area metric on $(M,g,G)$ is induced from the metric such that $G=G_g$, one finds $X=0$ such that the connection and curvature tensor on $\bigwedge^2TM$ coincide with the lifts induced by the Levi-Civita connection.

\section{Conclusion}\label{Conc}
Area metric manifolds present a true generalization of metric manifolds because only some area metrics are induced by a metric. The geometric structure of area metric manifolds is probed by minimal surfaces, the same way the structure of metric manifolds is probed by geodesics. However, a routine construction of covariant derivatives and tensors from the area metric is obstructed by the non-linear variety structure of the spaces of areas. We find that an area metric indeed only determines part of a connection, namely what we term a symmetric pre-connection. Although this structure can be extended to a connection by provision of an antisymmetric pre-connection, the latter is of no relevance to area geometry.

The equation for stationary surfaces is governed entirely by the symmetric pre-connection. In fact, minimal surfaces with tangent areas $\Omega$ are characterized by a vanishing mean curvature one-form,
\begin{equation}\label{seq}
\mathcal{D}_\Omega\Omega=0\,,
\end{equation}
where $\mathcal{D}_\Sigma\Omega$ is a derivative operator of areas acting on areas, constructed from the symmetric pre-connection. This identifies the operator $\mathcal{D}$ as the relevant differential geometric structure of area metric manifolds. Owing to the non-linearity of the space of areas,  $\mathcal{D}$ is merely $C^\infty$-homogeneous in $\Sigma$, and thus local.

String theory is most naturally interpreted from the point of view of area metric geometry. Indeed,
the Polyakov action, including target space metric $g$ and Neveu-Schwarz two-form~$B$, cannot be regarded as an area functional on a metric manifold. However, it is classically equivalent to a one-parameter family of surface area functionals based on area metrics $G_{g+\lambda^{-1}B}=G_g+\lambda^{-2}G_B$. Thus the associated symmetric pre-connection encodes the geometric structure $\mathcal{D}$ of generalized string backgrounds, and the string equation of motion takes the concise form (\ref{seq}).

This insight at the level of the fundamental string carries over to the string-induced effective action on a D-brane: D-branes are area metric manifolds $(M,G)$, but not metric manifolds. The Dirac-Born-Infeld action may be regarded as the volume form of an area metric, but not that of a metric. The area measure in this case is composed of the metric, the $B$-field and the gauge field strength on the brane as $G_g+\kappa G_{B+2\pi\alpha'F}$, which is discussed in more detail in the text. These results suggest that gravity on a D-brane should be described by the dynamics of an area metric rather than by the dynamics of the more restrictive structure of a metric. Following this line of thought will require the construction of scalar invariants from the operator $\mathcal{D}$, which is complicated due to the non-linearity of area spaces, but is currently under investigation.

The complications in the construction of such true area geometric invariants may be bypassed if additional structure is provided. On manifolds $(M,g,G)$ equipped with a metric and an independent area metric, scalar invariants can be constructed in standard fashion using the extension of the symmetric pre-connection to a connection. This is not entirely in the spirit of area geometry, but linked to ideas appearing in the context of higher gauge theories.

In conclusion, area metric manifolds can be regarded as a geometric imperative to consider strings: they only allow the discussion of stationary surfaces, not that of stationary worldlines. Moreover, they easily accommodate generalized backgrounds via area measures more general than those induced by a metric. Viewing string theory from this perspective motivates the study of the differential geometry of area metric manifolds. With the help of the differential operators constructed from the area metric we find that it is area metric manifolds which capture the structure of effective string backgrounds, rather than metric manifolds with additional fields.
For this reason we expect that further development of the geometry of area metric manifolds (which, for instance, could involve the development of appropriate notions of isometries, curvature and related invariants, or the investigation of holonomies) will shed light on effective dynamics appearing in string theory. In particular this should apply to the issue of gravity induced on D-branes, which is still largely unresolved. Here, area metric geometry provides constructional tools beyond those of the conventional differential geometry of metric manifolds.

\acknowledgments
The authors thank Andrei Starinets and Raffaele Punzi for valuable discussions. MNRW thanks the Perimeter Institute for their hospitality while this work was completed. He acknowledges financial support from the German Research Foundation DFG, the German Academic Exchange Service DAAD, and the European RTN program MRTN-CT-2004-503369.

\appendix
\section*{Appendix: Conventions}\label{conventions}
In this appendix we explain our conventions for the coordinate representations of objects and operations. Let $M$ be a $d$-dimensional smooth manifold and $\{e_a\}$ a basis of $TM$. This basis canonically induces the basis
\begin{equation}\label{basis1}
\{e_{a_1}\wedge\dots\wedge e_{a_k}\,|\,a_1<\dots<a_k\}
\end{equation}
of $\bigwedge^kTM$ which is of dimension $\left({}^d_k\right)$. Any section $\Omega$ of $\bigwedge^kTM$ can be expanded in this basis in the following form,
\begin{equation}
\Omega=\Omega^{a_1\dots a_k}e_{a_1}\wedge\dots\wedge e_{a_k}=\frac{1}{k!}\Omega^{ab\dots}e_a\wedge e_b\wedge\dots\,,
\end{equation}
which defines the components of $\Omega$. Note that we adhere to the convention that sums over numbered indices $a_1\dots a_k$ are ordered sums over  $a_1<\dots <a_k$. As an immediate consequence of this convention we find the components of exterior products of sections $\Omega\in \bigwedge^kTM$ and $\Sigma\in\bigwedge^lTM$:
\begin{equation}
(\Omega\wedge\Sigma)^{ab\dots}=\left({}^{k+l}_{\;\;k}\right)\Omega^{[ab\dots}\Sigma^{\dots]}\,.
\end{equation}
A basis $\{\epsilon^a\}$ of $T^*M$ induces the canonical basis $\{\epsilon^{a_1}\wedge\dots\wedge\epsilon^{a_k}\,|\,a_1<\dots <a_k\}$ of $\bigwedge^kT^*M$ which is dual to (\ref{basis1}) if $\{\epsilon^a\}$ is dual to $\{e_a\}$, in which case
\begin{equation}
\epsilon^{a_1}\wedge\dots\wedge\epsilon^{a_k}(e_{b_1}\wedge\dots\wedge e_{b_k})=k!\delta^{[a_1}_{[b_1}\dots \delta^{a_k]}_{b_k]}\,,
\end{equation}
where the right hand side are precisely the components of the identity on $\bigwedge^kTM$. We also define contraction symbols for $\Omega\in\bigwedge^{k+1}TM$ and $\omega\in\bigwedge^{l+1}T^*M$:\begin{subequations}
\begin{eqnarray}
\Omega\,\llcorner\,\omega &=& \textstyle{\frac{1}{k+1}\frac{1}{l+1}}\,\Omega^{a_1\dots a_kp}\omega_{pb_1\dots b_l}e_{a_1}\wedge\dots\wedge e_{a_k}\otimes\epsilon^{b_1}\wedge\dots\wedge\epsilon^{b_k}\,,\\
\omega\,\lrcorner\,\Omega &=& (-1)^{k+l}\,\Omega\,\llcorner\,\omega\,.
\end{eqnarray}
\end{subequations}
For k-forms $\Sigma\in\bigwedge^kT^*M$ we define the exterior derivative such that
\begin{equation}
d\Sigma=(k+1)\partial_{[a_1}\Sigma_{a_2\dots a_{k+1}]}\epsilon^{a_1}\wedge\dots\wedge\epsilon^{a_{k+1}}\,.
\end{equation}


\end{document}